\documentclass[twocolumn]{aastex62}

\usepackage{float}
\usepackage{multirow}
\makeatletter

\newcommand{\Rmnum}[1]{\expandafter\@slowromancap\romannumeral #1@}
\makeatother
\shortauthors{R. Wang et al.}

\begin{document}

\title{SPCANet: Stellar Parameters and Chemical Abundances Network for LAMOST-\Rmnum{2} Medium Resolution Survey}

\correspondingauthor{A-Li Luo*}
\email{lal@nao.cas.cn}

\author[0000-0001-6767-2395]{Rui Wang}
\affil{Key Laboratory of Optical Astronomy, National Astronomical Observatories, Chinese Academy of Sciences, Beijing 100101, China}
\affil{University of Chinese Academy of Sciences, Beijing 100049, China}

\author[0000-0001-7865-2648]{A-Li Luo*}
\affil{Key Laboratory of Optical Astronomy, National Astronomical Observatories, Chinese Academy of Sciences, Beijing 100101, China}
\affil{University of Chinese Academy of Sciences, Beijing 100049, China}
\affil{Department of Physics and Astronomy, University of Delaware, Newark, DE, 19716, USA}
\affil{Institute for Astronomical Science and School of Information Management, Dezhou University, Dezhou 253023, China}

\author{Jian-Jun Chen}
\affil{Key Laboratory of Optical Astronomy, National Astronomical Observatories, Chinese Academy of Sciences, Beijing 100101, China}

\author{Wen Hou}
\affil{Key Laboratory of Optical Astronomy, National Astronomical Observatories, Chinese Academy of Sciences, Beijing 100101, China}

\author[0000-0003-1454-1636]{Shuo Zhang}
\affil{Key Laboratory of Optical Astronomy, National Astronomical Observatories, Chinese Academy of Sciences, Beijing 100101, China}
\affil{University of Chinese Academy of Sciences, Beijing 100049, China}

\author{Yong-Heng Zhao}
\affil{Key Laboratory of Optical Astronomy, National Astronomical Observatories, Chinese Academy of Sciences, Beijing 100101, China}
\affil{University of Chinese Academy of Sciences, Beijing 100049, China}

\author{Xiang-Ru Li}
\affil{South China Normal University, Guangzhou 510631, China}

\author{Yong-Hui Hou}
\affil{Nanjing Institute of Astronomical Optics, \& Technology, National Astronomical Observatories, Chinese Academy of Sciences, Nanjing 210042,  China}
\affil{University of Chinese Academy of Sciences, Beijing 100049, China}

\author{LAMOST MRS Collaboration}
\affil{Department of Astronomy, Beijing Normal University, Beijing 100875, China}
\affil{Key Laboratory of Optical Astronomy, National Astronomical Observatories, Chinese Academy of Sciences, Beijing 100101, China}
\affil{Key Laboratory for Research in Galaxies and Cosmology, Shanghai Astronomical Observatory, Chinese Academy of Sciences, Shanghai 200030, China}
\affil{Key Laboratory for the Structure and Evolution of Celestial Objects, Yunnan Observatories, Chinese Academy of Sciences, Kunming 650216, China}
\affil{Kavli Institute for Astronomy and Astrophysics, Peking University, Beijing 100871, China}
\affil{School of Astronomy and Space Science, Nanjing University, Nanjing 210093, China}
\affil{Department of Physics, Hebei Normal University, Shijiazhuang 050024, China}
\affil{Institute for Astronomical Science and School of Information Management, Dezhou University, Dezhou 253023, China}
\affil{Department of Astronomy and Institute of Theoretical Physics and Astrophysics, Xiamen University,  Xiamen 361005, China}

\begin{abstract}
The fundamental stellar atmospheric parameters ($\textit{T}_{\text{eff}}$ and log $g$) and 13 chemical abundances are derived for medium-resolution spectroscopy from LAMOST Medium-Resolution Survey (MRS) data sets with a deep-learning method. The neural networks we designed, named as SPCANet, precisely map LAMOST MRS spectra to stellar parameters and chemical abundances. The stellar labels derived by SPCANet are with precisions of 119 K for $\textit{T}_{\text{eff}}$ and 0.17 dex for log $g$. The abundance precision of 11 elements including [C/H], [N/H], [O/H], [Mg/H], [Al/H], [Si/H], [S/H], [Ca/H], [Ti/H], [Cr/H], [Fe/H], and [Ni/H] are 0.06$\sim$0.12 dex,  while of [Cu/H] is 0.19 dex. These precisions can be reached even for spectra with signal-to-noise as low as 10. The results of SPCANet are consistent with those from other surveys such as APOGEE, GALAH and RAVE, and are also validated with the previous literature values including clusters and field stars. The catalog of the estimated parameters is available at \url{http://paperdata.china-vo.org/LAMOST/MRS_parameters_elements.csv}. 
\end{abstract}


\section{Introduction}           %% first-level sections will be auto-capitalized
\label{sect:intro}

Large scale spectroscopic surveys (e.g., SDSS/SEGUE: \citet{2009AJ....137.4377Y},  LAMOST/LEGUE: \citet{2015RAA....15.1095L}, SDSS/APOGEE: \citet{2017AJ....154...94M}, RAVE: \citet{2006AJ....132.1645S}, Gaia-ESO: \citet{2012Msngr.147...25G}, GALAH: \citet{2015MNRAS.449.2604D}, Gaia-RVS: \citet{2004MNRAS.354.1223K}) have produced huge amount of precious spectroscopic data for lifting the veil of the Milky Way. The spectra of these surveys cover optical to near infrared spectral bands with low, medium and high resolving power depending on their specific science goals. The main stellar parameters, including the effective temperature ($\textit{T}_{\text{eff}}$), the surface gravity (log $g$), chemical abundances and radial velocity (RV) are the major information derived from spectra and are valuable materials for both Galactic archaeology and stellar evolution history. 

LAMOST-\Rmnum{2} medium resolution (R$\sim$7500) spectroscopic survey (MRS) started running after LAMOST-\Rmnum{1}~\citep{2015RAA....15.1095L} obtained more than 9 million spectra during its first five-year regular survey with the low-resolution mode (R$\sim$1800). LAMOST-\Rmnum{2} MRS aims to obtain the most numbers of medium optical bands spectra for researchers. Six main scientific working groups are providing observation plans serving their scientific goals including Galactic archeology, time-domain astronomy, star formation, open cluster, nebulae, exo-planets etc. For details, we refer readers to read the paper (Liu et al. in preparation). 

Most of the spectroscopic surveys developed pipelines to estimate fundamental stellar parameters and some of the pipelines has the capability to obtain element abundances information. LAMOST is equipped with a stellar parameters pipeline~\citep[LASP;][]{2015RAA....15.1095L,2011RAA....11..924W} adapted from ULySS package~\citep{2009A&A...501.1269K}, SEGUE published a stellar parameter pipeline~\citep[SSPP;][]{2008AJ....136.2022L} involving multiple techniques, fitting observations with synthesis spectra grids, artificial neural networks and empirical relations methods, APOGEE also set up their stellar parameter and chemical abundances pipeline~\citep[ASPCAP;][]{2016AJ....151..144G,2018AJ....156..126J} parametrizing near-infrared spectra by minimizing $\chi^{2}$ between observations and theoretical spectra, and RAVE holds the pipeline ~\citep{2006AJ....132.1645S} which exploits a best-matched template to measure radial velocities and atmospheric parameters. 

The traditional methods of determining stellar parameters are mostly based on mining k-square between observation and synthetic model spectra or empirical spectral libraries to search for the best-fit reference spectra. These reference spectra are always subject to different kinds of defects. Synthetic model spectra come from the stellar photosphere models which depend on oversimplified physical assumptions in some condition that led to inconsistent with observed data. For the empirical spectral libraries, the parameter coverage space, spectral resolving power, and the wavelength coverage can not always meet the scientific requirements of each survey. For example, the ELODIE spectral library~\citep{2007astro.ph..3658P} lacks of K giants and sub-giants samples, and its wavelength coverage is limit within 4000-6800\AA . The wavelength coverage of the MILES library~\citep{2011A&A...532A..95F} spectra are wider than that of ELODIE, while its resolution is as low as 2000, which cannot be used as reference sets for medium resolution spectral surveys. 

For most of high resolution spectra data sets, their precise element information are extracted from the absorption lines based on the measurement of equivalent widths (EWs) or the computation of synthetic spectra\citep{2019ARA&A..57..571J}. However, for most medium resolution spectra, these methods do not play a desired role to precisely derive elemental abundances because of line blending. Recently with the successful application in dealing with large scale, multi-dimension data, the data driven methods, on the other hand, become an option on transferring stellar parameters and chemical abundances of high resolution spectra to low or medium resolution spectra. In this way, the number of stellar chemical abundances would be extremely expanded and the precision of stellar labels of lower resolution spectra can be improved to a high level.

Researchers made efforts on machine learning in stellar parameter estimation, such as \emph{The Cannon}~\citep{2015ApJ...808...16N,2016arXiv160303040C}, \emph{The Payne}~\citep{2019ApJ...879...69T}, \emph{StarNet}~\citep{2018MNRAS.475.2978F}, \emph{AstroNN}~\citep{2019MNRAS.483.3255L} and \emph{GSN}~\citep{2019PASP..131b4505R}, and most of them employ artificial neural networks for building regression map relationship. These methods depend on training and test sets usually called reference sets, and the more complete the parameter space covers, the more information can be obtained by the model training. The best reference sets used for machine learning algorithms are from collections of so called standard stars with reliable stellar labels derived from high resolution spectra or more precise estimations. However, the numbers of the stars in these collections are always the bottleneck which leads to sparse distribution in the whole parameter space. For example, APOGEE DR14~\citep{2018AJ....156..125H} released parameters and elemental abundances derived by ASPCAP~\citep{2016AJ....151..144G,2018AJ....156..126J}, but are only reliable for giants. Gaia-ESO~\citep{2012Msngr.147...25G} and GALAH~\citep{2015MNRAS.449.2604D} released many high resolution spectra, but the corresponding stars all locates on the southern hemisphere so that very few common stars with surveys on the northern hemisphere can be used for labels transfer learning.

The theoretical spectral grid is an alternative. \emph{The Payne} \citep{2019ApJ...879...69T} was trained using the theoretical spectra based on Kurucz models with improved line lists (Cargile et al. in preparation).  This neural network performs better than the quadratic model (\emph{The Cannon}) in characterizing the non-linear and complex relationship between stellar labels and near infrared spectral fluxes. It was applied in estimating stellar parameters and elemental abundances for $\sim$230,000 near infrared spectra of APOGEE DR14 including both giants and dwarfs. \emph{The Payne} results without any calibration required show very good performance, which are much more precise than ASPCAP published values calibrated based on empirical relationships and some information from star clusters. Thus the star labels of APOGEE objects derived through \emph{The Payne} provide us a precious reference sets for training data-driven model. 

In this paper, we design a convolutional neural network model, named as SPCANet, to map LAMOST-\Rmnum{2} MRS spectra to star labels rather than reproduce spectra given star labels like \emph{The Payne} does. The SPCANet relies on on the advantage of neural networks that initial feature selection is not required comparing with other machine learning algorithms. We cross-match LAMOST-\Rmnum{2} MRS with \emph{The Payne} catalog and get 12,433 common stars corresponding 98,612 MRS spectra as the reference set for SPCANet, which means that all spectra in the training and testing sets are real LAMOST MSR spectra in two specific optical windows, and corresponding star labels are from \emph{The Payne}. 

This paper is organized as follows. Sect.\ref{sect:data} briefly introduces the LAMOST-\Rmnum{2} Medium Resolution Survey and corresponding data, as well as the reference data sets obtained for training and testing SPCANet. Sect.\ref{sect:methods} focuses on the design and the training process of SPCANet. Sect.\ref{sect:results} highlights the application result of SPCANet in determining the stellar atmospheric parameters and elemental abundances.  Sect.\ref{sect:discussion} discusses some challenges associated with this research followed by a summary in Sect.\ref{sect:summary}.  

\section{Data}
\label{sect:data}

The datasets studied in this work consist of two parts: LAMOST-\Rmnum{2} MRS spectra and the reference catalog of stellar parameters (effective temperature \textit{T}$_{\text{eff}}$ and surface gravity log \textit{g}) and 13 elemental abundances ([Fe/H], [C/H], [N/H], [O/H], [Mg/H], [Al/H], [Si/H], [S/H], [Ca/H], [Ti/H], [Cr/H], [Ni/H], [Cu/H]) from \citet[][APOGEE-\emph{Payne} catalog]{2019ApJ...879...69T} used for training and testing the data-driven model. 

\subsection{LAMOST MRS Observations}

\subsubsection{Observations}
The Large sky Area Multi-Object fiber Spectroscopy Telescope (LAMOST) is a reflecting Schmidt telescope combining a large aperture and a large field of view, both of which backup a highly-multiplexed spectroscopic system. It locates in Xinglong Observatory, Hebei province, China. The focal surface of LAMOST is circular with a diameter of 1.75 meters ($\sim 5^{\circ}$), and 4000 fibers are almost evenly distributed over it. Each of the fibers can be moved with two degrees of freedom by two motors. The light of 4000 objects observed simultaneously is transmitted to 16 spectrographs through fibers and recorded by 32 4K*4K charge-coupled devices (CCD). 

\begin{figure}[htbp]
\centering
\includegraphics[width=0.5\textwidth, angle=0]{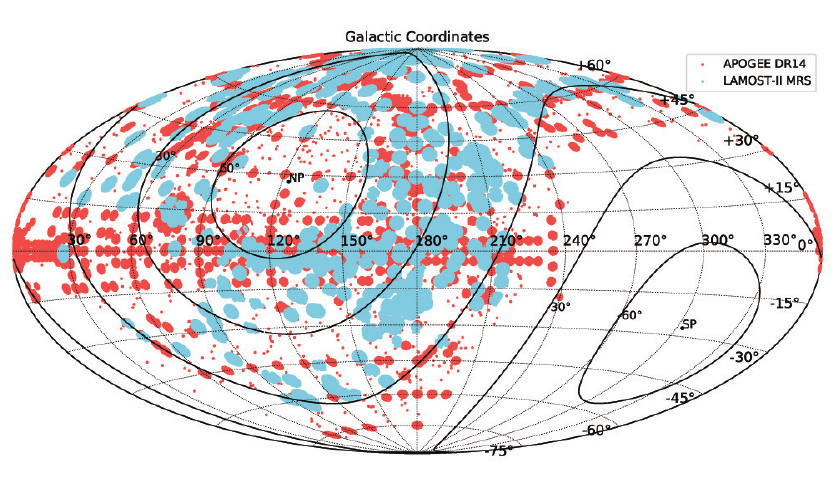}
\caption{Distribution of the Galactic coordinates of LAMOST-\Rmnum{2}  MRS DR7 (blue) and APOGEE DR14 (red). NP and SP in the figure refer to the North Pole and the South Pole of celestial coordinates respectively.}
\label{Figure1}
\end{figure}

\begin{figure}[htbp]
\centering
\includegraphics[width=0.5\textwidth, angle=0]{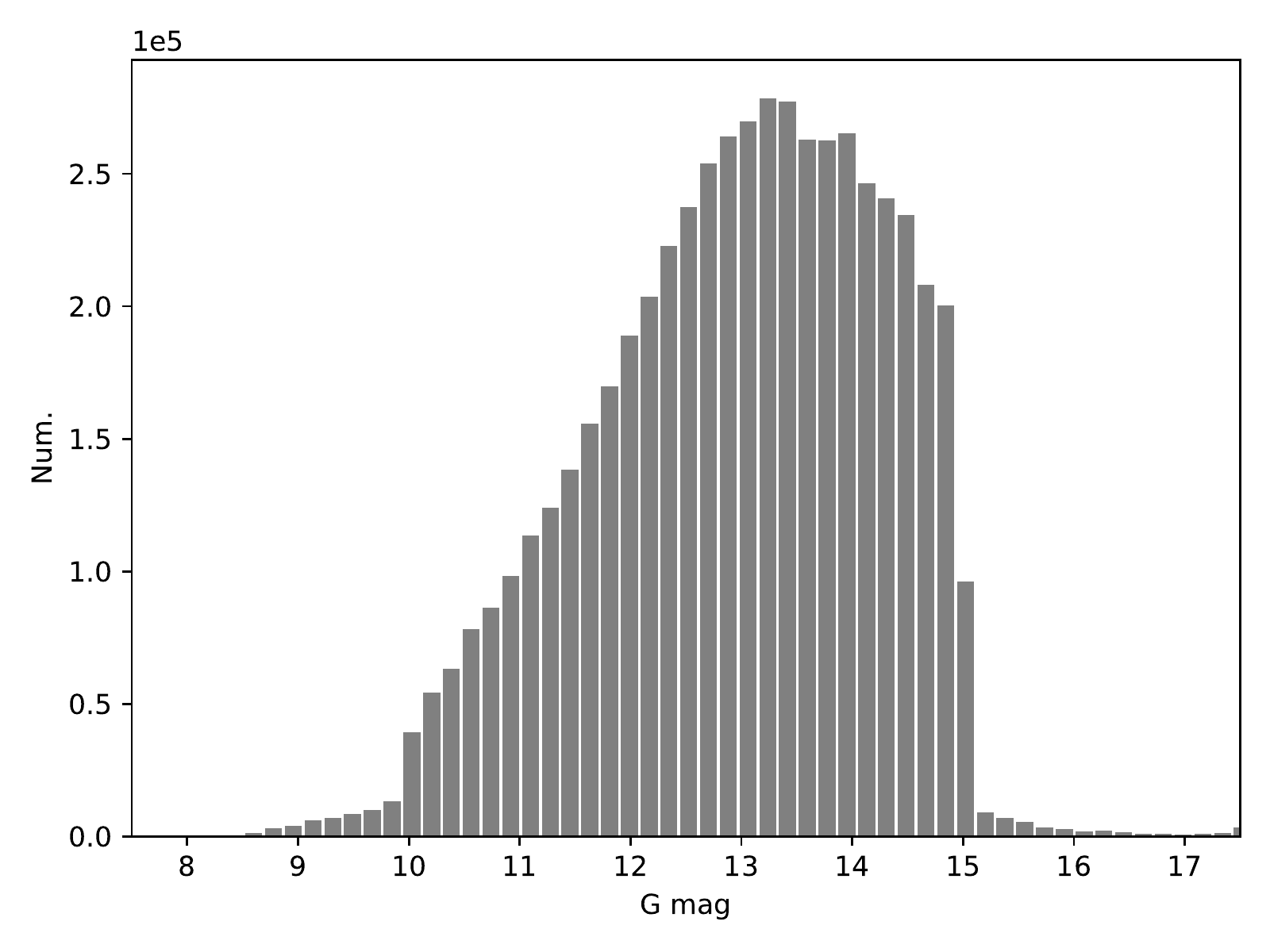}
\caption{Distribution of the G magnitude employed from the Gaia DR2 photometric catalog of LAMOST MRS DR7.}
\label{Figure2}
\end{figure}

\begin{figure}[htbp]
\centering
\includegraphics[width=0.5\textwidth, angle=0]{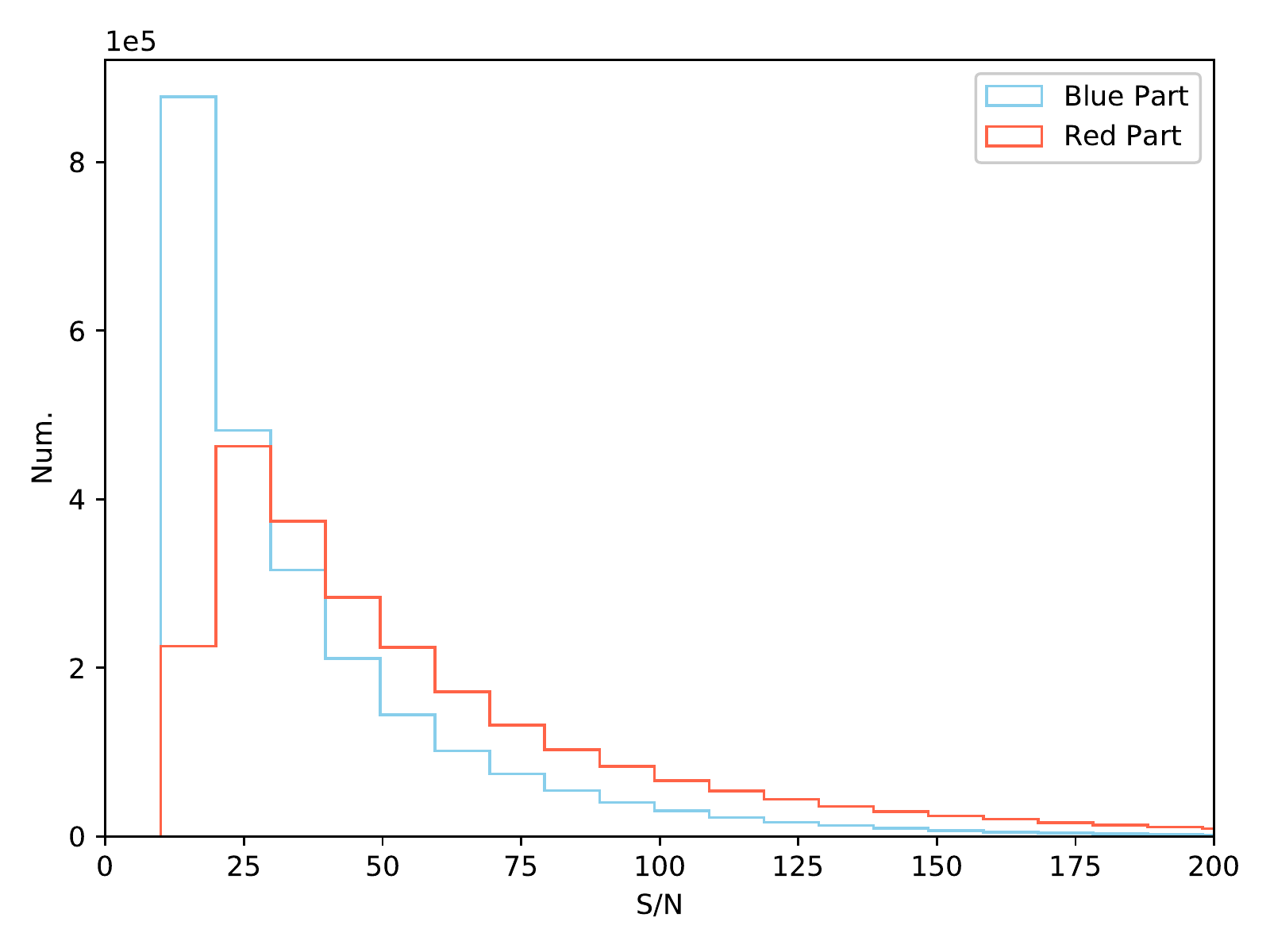}
\caption{Distribution of the signal-to-noise (S/N) of LAMOST MRS DR7 with S/N higher than 10 for both blue and red parts, blue color stands for blue parts, and red for red parts. }
\label{Figure3}
\end{figure}

LAMOST spectrograph has two resolving modes, which are the low-resolution mode of R$\sim$1800 and the medium-resolution mode of R$\sim$7500 respectively. The medium-resolution survey, LAMOST-\Rmnum{2} MRS, began on 1st Sep 2017 after the first five-year regular low-resolution survey ~\citep{ 2015RAA....15.1095L}. The wavelength coverage of each MRS spectrum is consists of two parts: the blue part (4950-5350 \AA) and the red part (6300-6800 \AA). LAMOST DR7 internally released 5,635,640 medium resolution spectra, 2,426,237 of which are with signal-to-noise (S/N) higher than 10 for both blue and red part. Here, S/N is defined as an average value in a wavelength band and indicates the S/N \emph{per pixel}. The ``footprints", the distribution of the $\emph{G}_{\text{mag}}$ by cross-matching Gaia DR2 photometic catalog and the distribution of signal-to-noise (S/N) of LAMOST-\Rmnum{2} MRS observations are shown in Fig.\ref{Figure1}, Fig.\ref{Figure2} and \ref{Figure3} respectively. We can see that most of MRS observations are concentrated in the $\emph{G}_{\text{mag}}$ range 10 to 15 mag and S/N of red parts are slightly higher than that of blue parts.

\subsubsection{Data Reduction}

\begin{figure*}[htbp]
\centering
\includegraphics[width=\textwidth, angle=0]{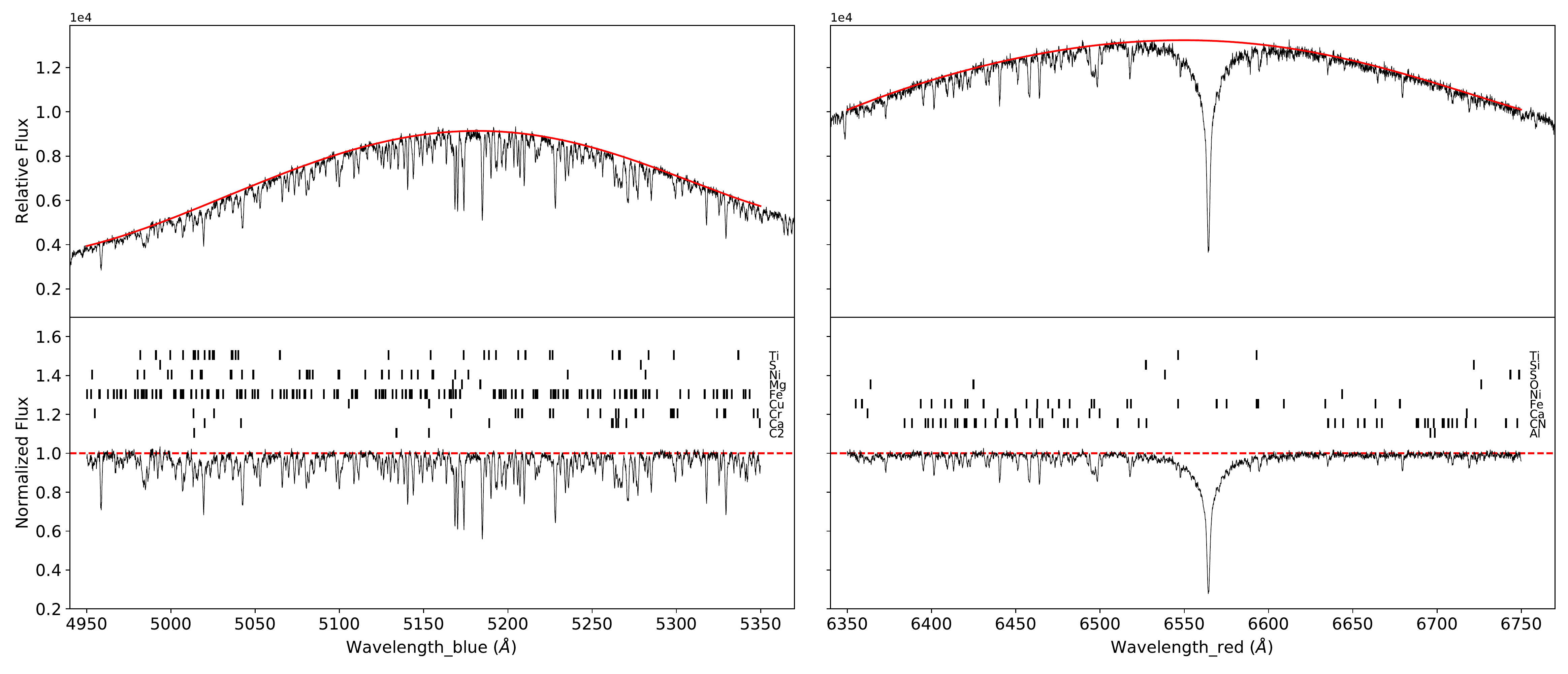}
\caption{An example of a LAMOST MRS spectrum: relative flux (top) and continuum-normalized flux (bottom) for the blue (left) and red (right) part. The red solid curves in the top panels are the pseudo-continuum fitted by the method as \citet{2008AJ....136.2022L} iteratively rejecting the points which lie 1$\sigma$ below and 4$\sigma$ above the fitted function to remove  strong absorption lines such as Balmer lines, the pseudo-continuum is obtained from a 4th-order polynomial. The red dotted lines in the bottom panels are plotted as a reference. The small short black lines in the bottom panels show the metal absorption lines, some of which are weak and blended in the medium resolution. The metallic lines in the blue part are more densely distributed than those in the red part.}
\label{Figure4}
\end{figure*}

LAMOST-\Rmnum{2} MRS spectra are processed using the same standard pipeline as low-resolution spectra used~\citep{2015RAA....15.1095L}, including the steps of bias subtraction, fiber tracing, fiber flat fielding, wavelength calibration, and sky subtraction etc.). One change of wavelength calibration for medium-resolution spectra from low-resolution spectra is using new Arc lamps (Th-Ar and Sc) for blue and red respectively instead of the old ones (Hg-Cr). \citet{2019ApJS..244...27W} made efforts to precisely measure radial velocities (RV) for objects of LAMOST-\Rmnum{2} MRS by cross-correlation with more than 2000 Kurucz model spectra~\citep{1993sssp.book.....K,2004astro.ph..5087C}. Using their method, for the spectra with signal-to-noise ratio(S/N) higher than 10, the precision of RVs can achieved as high as 1.36 $\text{km}~\text{s}^{-1}$. All the spectra are shifted to rest-frame according to the RV measured through above method, and spectra with S/N$>$10 are selected to measure their stellar parameters and chemical abundances. The spectra shifted to rest-frame are resampled in a step of 0.1 \AA~within two fixed wavelength coverage: 4950-5350 \AA~for the blue part and 6350-6750 \AA~for the red part. Here, we sample 4000 'pixels' in both the blue and the red part to keep two homogeneous inputs for the two parallel branches of our neural network. For each part, the spectrum is normalized after obtaining a pseudo-continuum. The continuum fit is same as \citet{2008AJ....136.2022L}: iteratively rejecting the points which lie 1$\sigma$ below and 4$\sigma$ above the fitted function to remove  strong absorption lines such as Balmer lines, the pseudo-continuum is obtained from a 4th-order polynomial. An example of LAMOST-\Rmnum{2} MRS raw spectra and corresponding continuum-normalized spectra are both shown in Fig.\ref{Figure4}. 

\subsection{Reference set}

The Apache Point Observatory Galactic Evolution Experiment~\citep[APOGEE;][]{2015AJ....150..148H, 2017AJ....154...94M} is a median-high resolution (R$\sim$22,500) spectroscopic survey in the near-infrared spectral range (H band, $\lambda = 15700$ to $17500$ \AA).  The Data Release 13 and 14 (DR13, DR14) of APOGEE was described in detail in the paper authored by ~\citet{2018AJ....156..125H}, in which stellar parameter results from a data-driven technique, \emph{The Cannon}, were also carefully discussed. \citet{2019ApJ...879...69T} proposed a Neural Network, \emph{The Payne}, to estimate effective temperature, surface gravity, and 15 element abundances for both giants and dwarfs in APOGEE DR14. \emph{The Payne} was trained through Kurucz model based synthetic grid with state-of-art line lists (Cargile et al, in preparation), and the results of the well trained \emph{The Payne} show high accuracy and precision without calibration and made up for the lack of dwarf stars in the official parameter catalog produced through APOGEE Stellar Parameters and Chemical Abundances Pipeline~\citep[ASPCAP;][]{2016AJ....151..144G,2018AJ....156..126J}.

\citet{2019ApJ...879...69T} derived stellar labels for totally 222,707 stars in the parameter ranges 3050 K $<\textit{T}_{\text{eff}}< $ 7950 K, 0 $<$ log \textit{g} $<$ 5 and -1.45 $<$ [Fe/H] $<$ 0.45, and excluding dwarfs with \textit{T}$_{\text{eff}} <$ 4000 K which are considered unreliable. APOGEE-\emph{Payne} catalog achieved an accuracy of 30 K for \textit{T}$_{\text{eff}}$, 0.05 dex for log \textit{g} and better than 0.05 dex for all the 15 elemental abundances (C, N, O, Mg, Al, Si, S, K, Ca, Ti, Cr, Mn, Fe, Ni, and Cu). We cross-match LAMOST-\Rmnum{2} MRS DR7 which S/N are higher than 10 with APOGEE-\emph{Payne} catalog and obtained 12,433 common stars corresponding 98,612 LAMOST-\Rmnum{2} MRS spectra after limiting the APOGEE-\emph{Payne} catalog \ ``quality\_flag" as \ ``good". In the wavelength window of LAMOST spectra, we examine each elemental feature and chose 13 elements (C, N, O, Mg, Al, Si, S, Ca, Ti, Cr, Fe, Ni and Cu) as our objective elements for measuring abundances. 

\section{Method}
\label{sect:methods}

Artificial neural network (ANN) methods were firstly adopted to determine stellar atmospheric parameters by ~\citet{1997MNRAS.292..157B}, and rejuvenated recently because of development of new training techniques and hardware. Inspired by the successful application of convolutional neural networks (CNN) to APOGEE spectra~\citep{2018MNRAS.475.2978F,2019MNRAS.483.3255L}, we design a specific CNN  structure for transferring stellar labels from APOGEE-\emph{payne} catalog to LAMOST-\Rmnum{2}  MRS spectra. 

\subsection{SPCANet: Stellar Parameters and Chemical Abundances networks}

ANNs work as simulating human and animal neuronal responses by mathematically connecting nodes of input, hidden and output layers. Many functional layers and connection form are developed with different effects, such as dense layers work by linear connection and non-linear activation to build a complicated non-linear function mapping, as: 
\\
\\
$y_j^{l+1}=g( \sum\limits_{i=0}^n (w_{ij}^l y_{ij}^l+b_{j}^l))$
\\
\\
where $g$ is an activation function, $w_{ij}$ is the weight representing the connection of the node $i$ of layer $l$ and node $j$ of layer $l+1$, and $b_{ij}^l$ is the bias of the node $i$ of layer $l$. 

\begin{figure}[htbp]
\centering
\includegraphics[width=0.5\textwidth, angle=0]{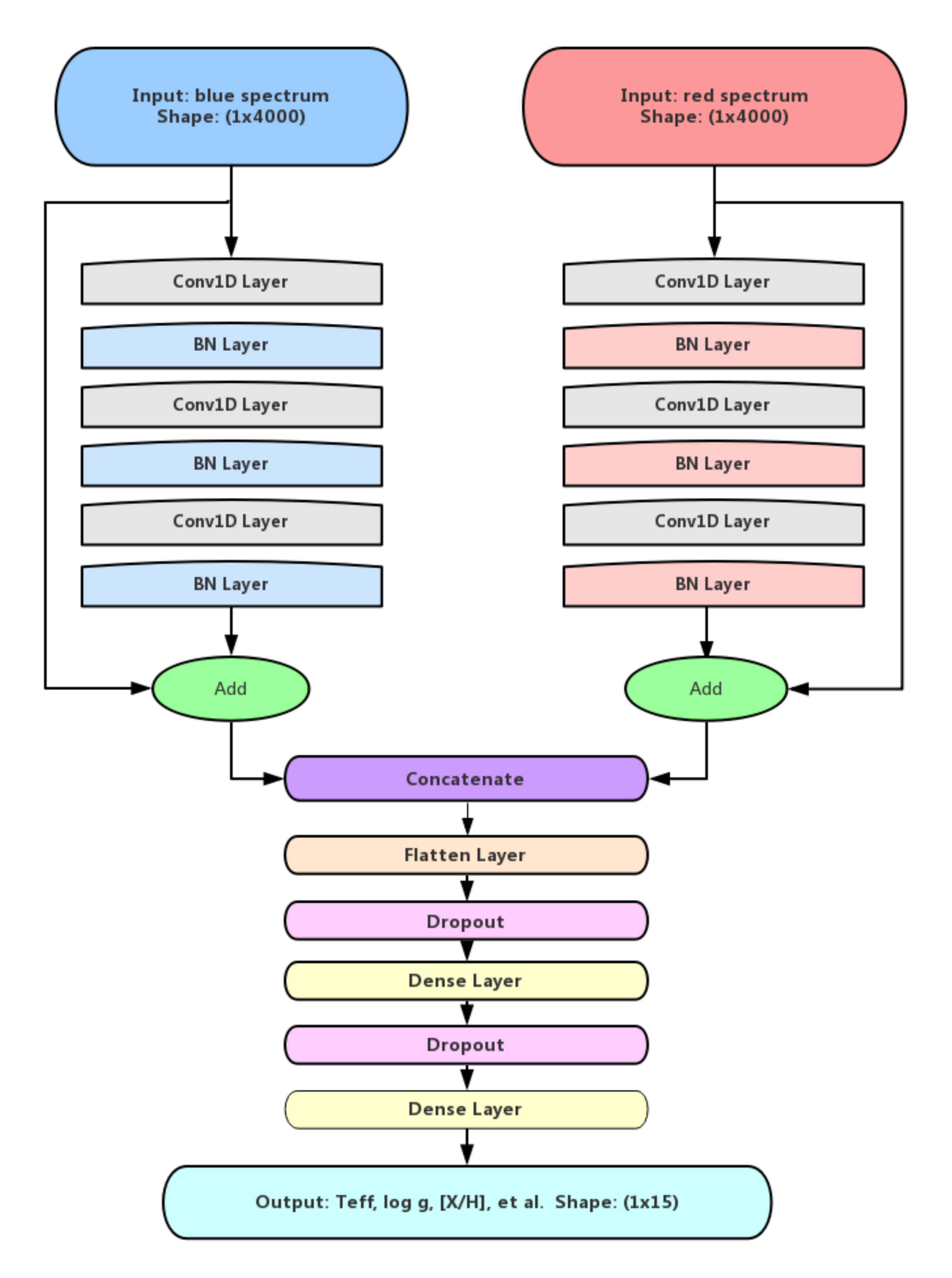}
\caption{SPCANet: convolutional neural networks designed for the estimation of stellar parameters and chemical abundances of LAMOST MRS observations. The inputs of two parallel branches of the networks are separately the blue/red continuum-normalized re-sampled (to fixed 4000 dimensions) spectra. After extracting features by three convolutional layers (Conv1D Layer) and three batch-normalization layers (BN Layer), their feature layers are stitched together after adding the input layers and then regressed to the output layer through two fully-connected layers (Dense Layer). The drop-out layers are employed among the fully-connected layers to overcome over-fitting. Finally, the output layer produce 15 stellar labels: $\textit{T}_{\text{eff}}$, log $g$ and 13 chemical abundances ([X/H]).}
\label{Figure5}
\end{figure}

Since each LAMOST-\Rmnum{2} MRS spectrum consists of two separate parts, inspired by the ResNet~\citep{2015arXiv151203385H}, we design a semi-parallel structure for SPCANet shown in Fig.~\ref{Figure5}. The SPCANet has two sets of double convolutional layers connected to the separate input layer for the blue and red part respectively. The output of the third convolutional layers for two branches add their input layers and then connect to the concatenate layer. Features from both branches finally map together to the output layer of stellar labels by two dense layers, and the Dropout~\citep{2012arXiv1207.0580H} steps are employed twice to avoid over-fitting among the dense layers. 

\subsection{Training and Testing of the model}

Deep networks~\citep{2015Natur.521..436L} always hold a huge amount of weights and hyper-parameters to be trained and fine-tuned by minimization of the loss function with the gradient descent algorithm, such as Adaptive Moment Estimation method~\citep[Adam;][]{2014arXiv1412.6980K}. Most of deep networks successfully work depending on huge amount of labeled samples which can constrain the model weights well, or on regularization and data augment technique to overcome the lack of enough labeled sets. The LAMOST-APOGEE\emph{payne} common stars are not sufficient for training a complex network structure, so we retain all the multi-epoch spectra of these stars from repeat observations which can be considered as independent observations because of different observational condition. Although the repeated observation do not change as a function of stellar parameters, the random errors from the different observing, instrumental, data reduction conditions would improve the generalization ability of the model. Because the randomicity of the errors would force the model learning the information from stars rather than from the noise. In this way, we have totally 98,612 LAMOST MRS spectra with \emph{payne} stellar labels and randomly divide them into the training set, the test set and the cross-validation set according to the ratio of 6:2:1. 

\begin{figure}[htbp]
\centering
\includegraphics[width=0.5\textwidth, angle=0]{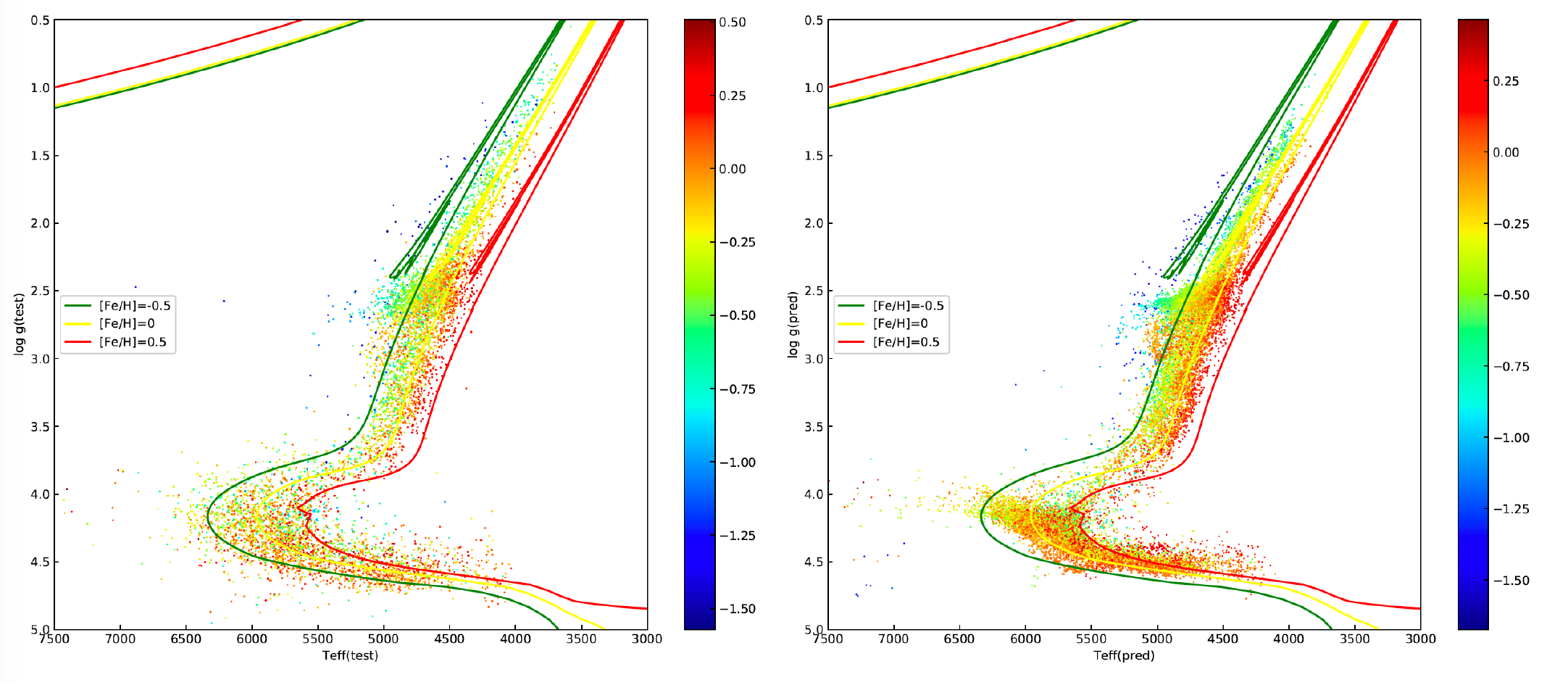}
\caption{Distribution of the test set (left panel) and the corresponding prediction by SPCANet model (right panel)l on \textit{T}$_{\text{eff}}$- log \textit{g} plane, color-coded by [Fe/H]. The over-plotted isochrones are employed from MIST stellar evolution assuming a stellar age of 7 Gyr and metallicities [Fe/H] of -0.5 (green), 0 (yellow) and 0.5 (red), respectively.}
\label{Figure6}
\end{figure}

\begin{figure*}[htbp]
\centering
\includegraphics[width=\textwidth, angle=0]{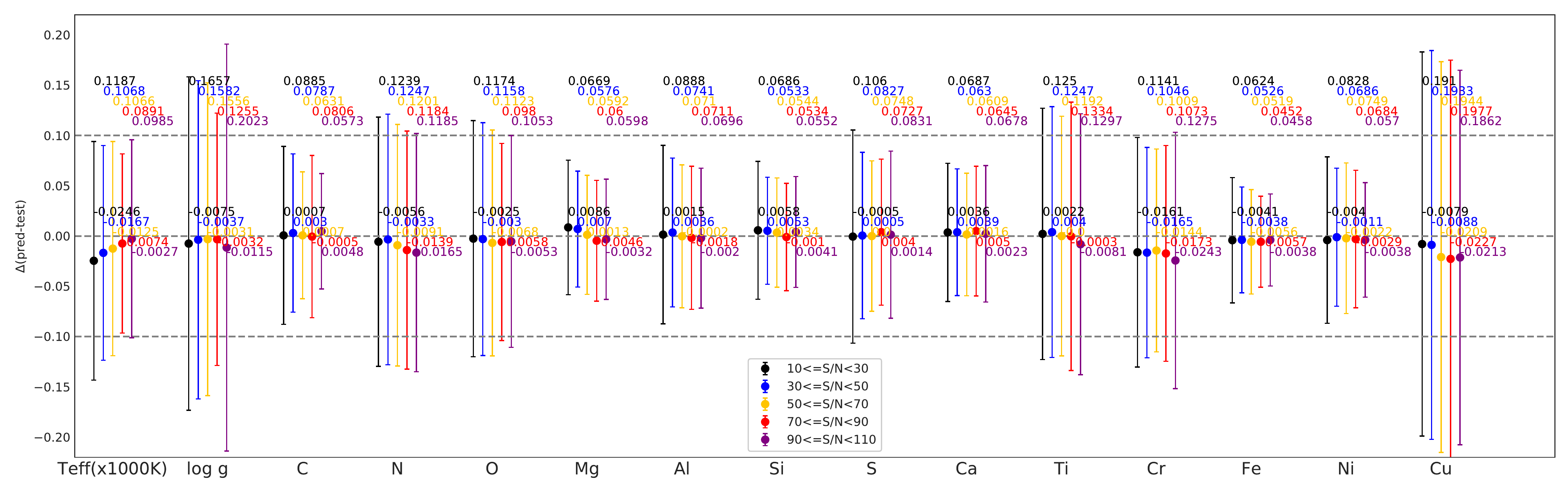}
\caption{Bias and 1-$\sigma$ errors between model predictions and the "true" values of stellar parameters and chemical abundances for SPCANet model test sets in different S/N intervals.}
\label{Figure7}
\end{figure*}

Fig.~\ref{Figure6} shows the distribution of the test set and the corresponding predictions by SPCANet model on \textit{T}$_{\text{eff}}$- log \textit{g} plane. We can see that SPCANet excellently reproduces stellar parameters which fit MIST isochrones~\citep{2016ApJ...823..102C,2016ApJS..222....8D} very well at the stellar age of 7 Gyr. Fig.~\ref{Figure7} compares the predictions by SPCANet model and their corresponding true stellar labels for about 20000 stars in the test sample in different S/N bins. For most stellar labels, the accuracy varies less with the S/N$_{\text{blue}}$ except \textit{T}$_{\text{eff}}$, which shows more accurate at higher S/N level. The precision decreases with the S/N$_{\text{blue}}$ increasing for several stellar labels while others are not, such as Ca, Ti and Cu, for which the precision show stable with S/N. The accuracy of \textit{T}$_{\text{eff}}$ and log \textit{g} derived from SPCANet model is 24.60 K and 0.0075 dex respectively, and the accuracy of 13 elemental abundances in the order of C, N, O, Mg, Al, Si, S, Ca, Ti, Cr, Fe, Ni and Cu is 0.0007 dex, 0.0056 dex, 0.0025 dex, 0.0086 dex, 0.0015 dex, 0.0058 dex, 0.0005 dex, 0.0036 dex, 0.0022 dex, 0.0161 dex, 0.0041 dex, 0.0040 dex and 0.0079 dex with S/N in [10,30] interval. In addition, the precision of \textit{T}$_{\text{eff}}$ and log \textit{g} from the model is 118.7 K, 0.1657 dex, and the precision of 13 elemental abundances in the same order is 0.0885 dex, 0.1239 dex, 0.1174 dex, 0.0669 dex, 0.0888 dex, 0.0686 dex, 0.1060 dex, 0.0687 dex, 0.1250 dex, 0.1141 dex, 0.0624 dex, 0.0828 dex, and 0.1910 dex. Most of the elements achieved 0.1 dex precision with S/N higher than 10, except N, O, Ti, Cr and Cu which have a little larger errors because their features are weak in most of the MRS spectra. 

\section{Results}
\label{sect:results}

\subsection{Predictions for LAMOST MRS spectra}
\begin{figure}[htbp]
\centering
\includegraphics[width=0.5\textwidth, angle=0]{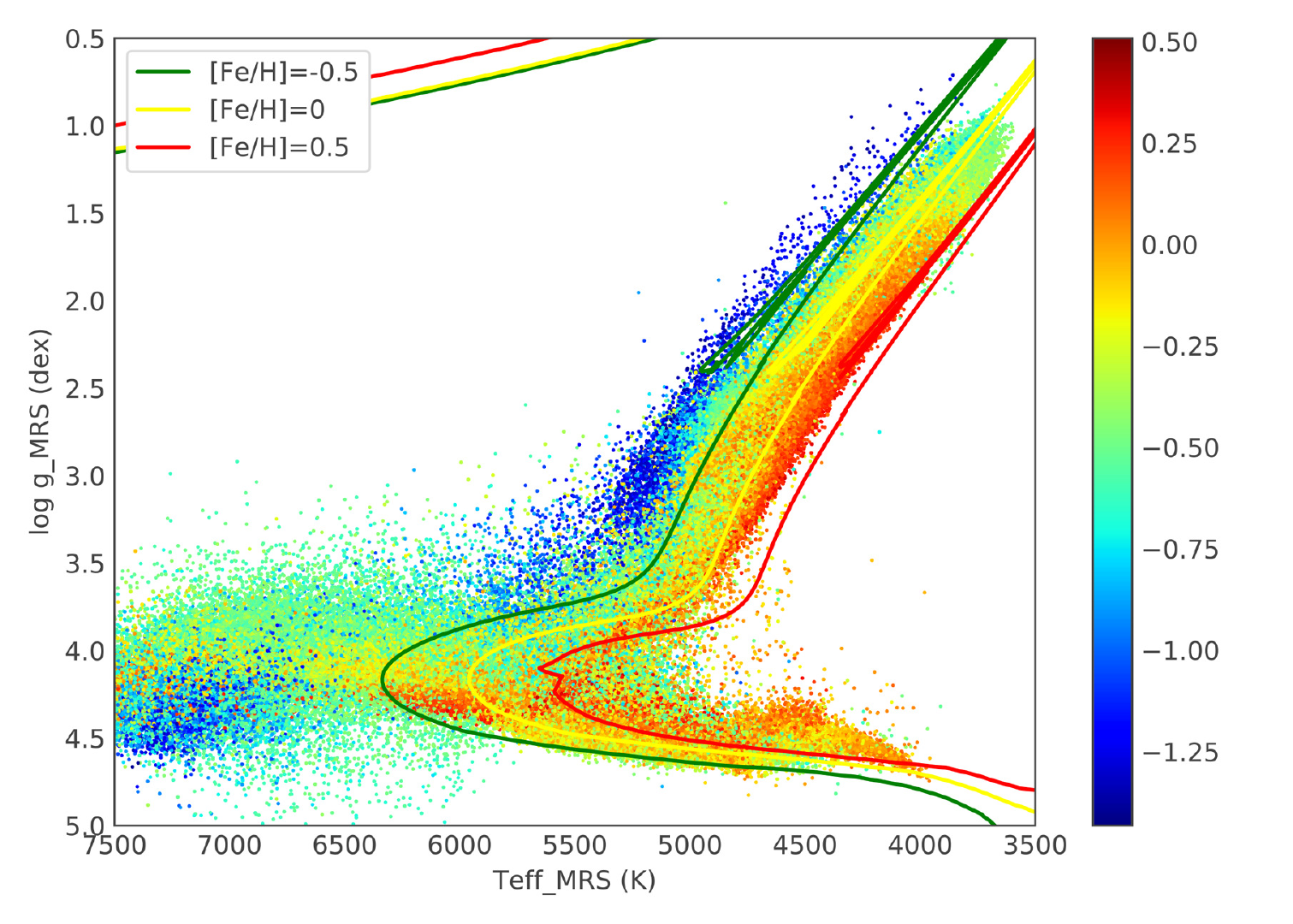}
\caption{Distribution of LAMOST MRS stellar parameters predicted by SPCANet on \textit{T}$_{\text{eff}}$-log \textit{g} panel, color-coded by [Fe/H]. The over-plotted isochrones are employed from MIST stellar evolution tracks with the stellar age of 7 Gyr with [Fe/H] of -0.5 dex, 0 and 0.5 dex.}
\label{Figure8}
\end{figure}

After training and testing, the SPCANet model is applied to estimate stellar parameters and chemical abundances for LAMOST-\Rmnum{2} DR7 MRS spectra. Pretreatment for all spectra are the same as that for the training set, including the wavelength shifted to rest-frame, the continuum normalized, fluxes re-binned. In general, we process and measure 2.4 million spectra and produce a catalog including stellar parameters and chemical abundances. Based on the range of the label of the training set we used, we excluded the targets with estimated temperatures above 8000 K or below 3500 K which are considered unreliable. To ensure the robustness of results, we also keep another alternate SPCANet model for examination. The alternate model which is trained on a training set differ from that of the formal model but perform comparably to the formal one. Only the stellar parameter results with little difference  (\textit{T}$_{\text{eff}}$ of 120 K, log \textit{g} of 0.16 dex and [Fe/H] of 0.06 dex) between two models' predictions would be kept in the final catalog.

\begin{figure*}[htbp]
\centering
\includegraphics[width=\textwidth, angle=0]{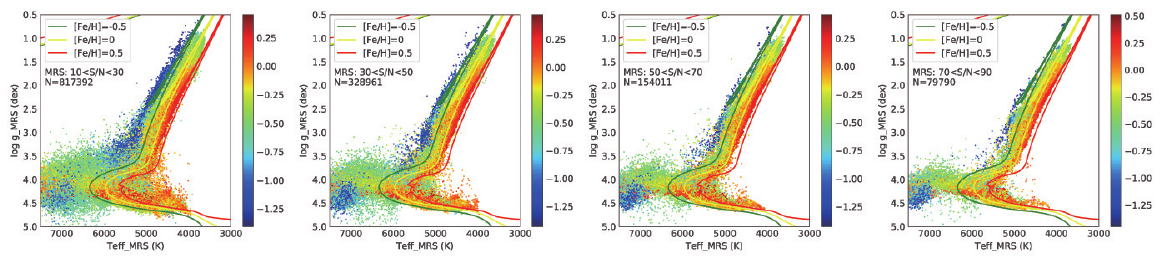}
\caption{Distribution of LAMOST MRS stellar parameters of stars in four S/N intervals, color-coded by [Fe/H].The over-plotted isochrones are employed from MIST stellar evolution tracks with the stellar age of 7 Gyr with [Fe/H] of -0.5 dex, 0 and 0.5 dex.}
\label{Figure9}
\end{figure*}

Distribution of LAMOST-\Rmnum{2} MRS in the \textit{T}$_{\text{eff}}$-log \textit{g} panel color-coded by [Fe/H] are shown in Fig.~\ref{Figure8}, and the over-plotted isochrones are employed from MIST stellar evolution tracks with the stellar age of 7 Gyr. The hot end of the main-sequence at (\textit{T}$_{\text{eff}}$, log \textit{g}) $\sim$ (7200 K, 4.5 dex) is populated by stars hotter than F, whose spectra in the MRS red part are dominated by strong Balmer lines (H$\alpha$ line), while their spectra in the MRS blue part lack features of metallic absorption lines. This means that less and degenerate information on the stellar parameters is provided by the spectra. In addition, high rotation, characteristic of hot stars, can add extra ``blurring" to the already relatively featureless spectra. The scarcity of training examples in this region of the parameter space also increase the error of the prediction for hot stars. Some stars at the cool end of the main-sequence (\textit{T}$_{\text{eff}}\sim[4000~\text{K}, 4500~\text{K}]$) display lower-than-expected surface gravitates. These may due to both the intrinsic complexity of their spectra and the scarcity of training examples. Fig.~\ref{Figure9} shows how the distribution varies with S/N$_{\text{blue}}$. The shape of the distribution get more cleaner for higher S/N$_{\text{blue}}$ decreasing the number of stars in the hot subgiants region. However, even in highest S/N$_{\text{blue}}$ interval, the diagram still shows the presence of metal-poor main-sequence stars with temperature higher than 7000 K. Researchers should be cautious when using the stellar labels with both \textit{T}$_{\text{eff}}$ higher than 7000 K and [Fe/H] lower than -1.0 dex.

\begin{figure*}[htbp]
\centering
\includegraphics[width=\textwidth, angle=0]{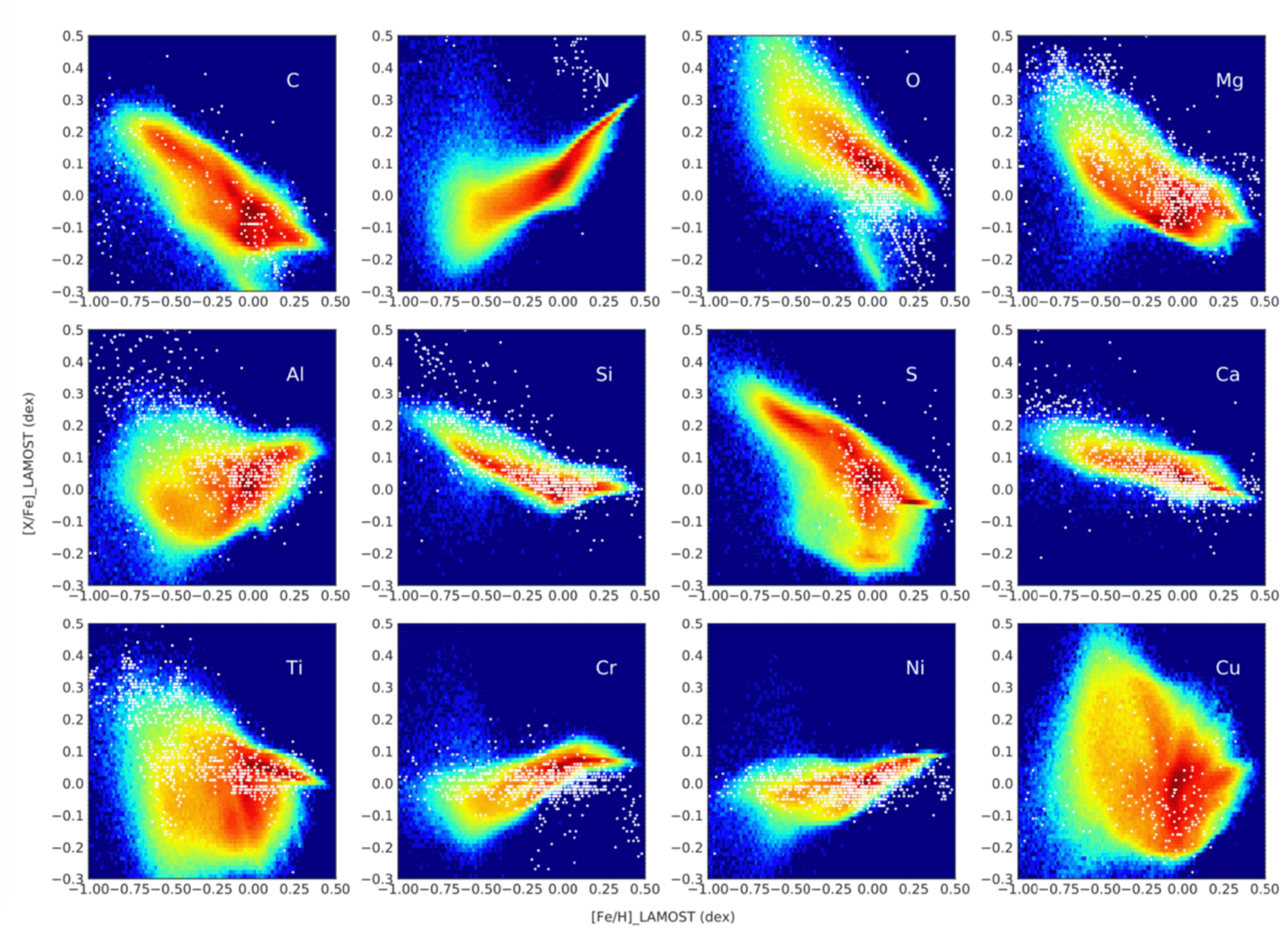}
\caption{Density distribution of elemental abundances with respect to [Fe/H] for LAMOST MRS dwarfs. The white scatters are values from the previous literature~\citep{2019A&A...622A.191M,2018AJ....156..142D,2019A&A...628A..54K,2017A&A...603A...2M,2016AJ....151...49K,2015A&A...573A..55T,2011ARep...55..689M,2014A&A...562A..71B,2016ApJ...833..225Z,2016A&A...593A..65N}.}
\label{Figure10}
\end{figure*}

\begin{figure*}[htbp]
\centering
\includegraphics[width=\textwidth, angle=0]{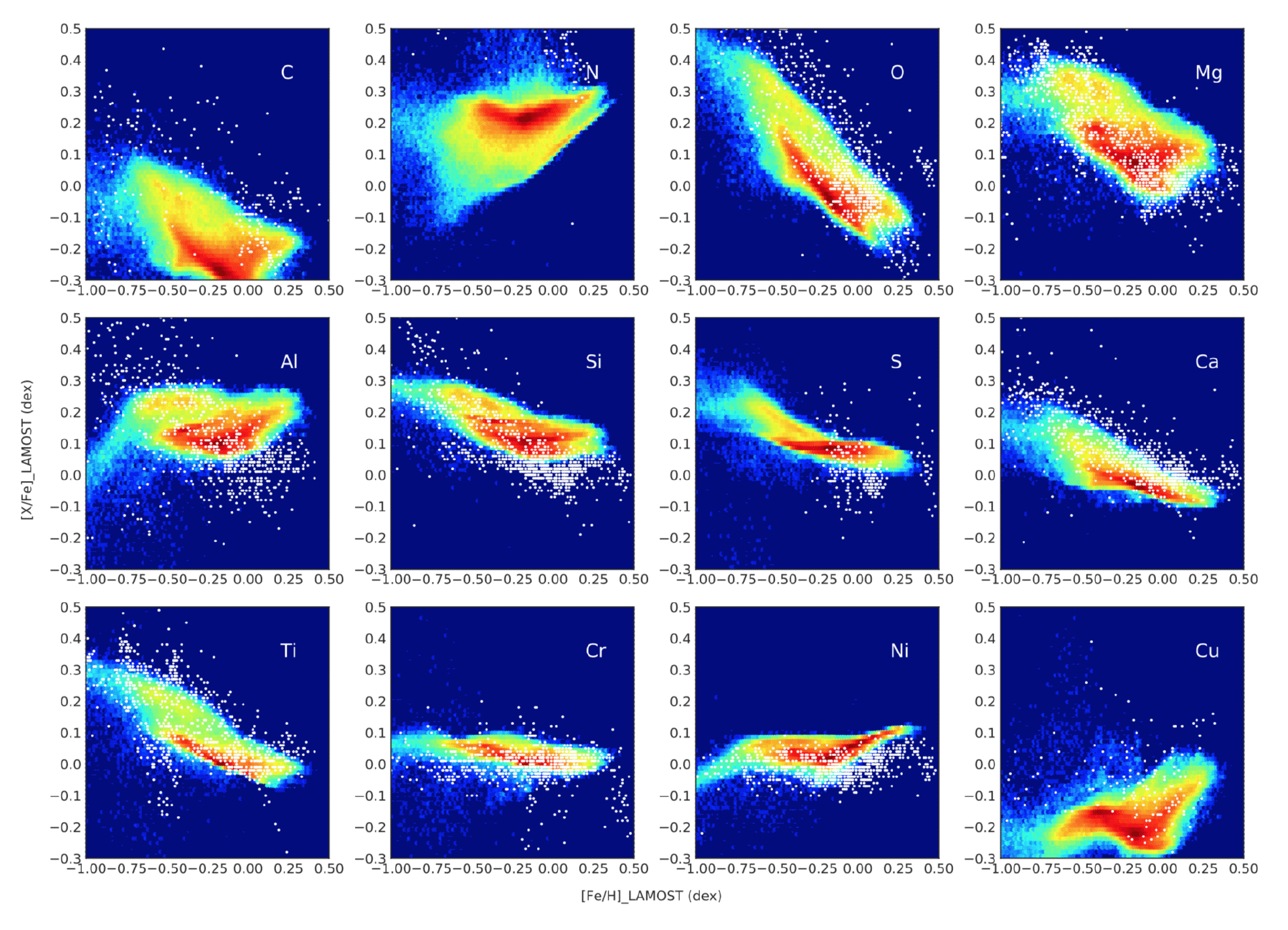}
\caption{Same as Fig.\ref{Figure10} but density distribution of elemental abundances for LAMOST MRS giants. The white scatters are values from the previous literature~\citep{2019A&A...622A.191M,2018AJ....156..142D,2019A&A...628A..54K,2017A&A...603A...2M,2016AJ....151...49K,2015A&A...573A..55T,2011ARep...55..689M,2014A&A...562A..71B,2016ApJ...833..225Z,2016A&A...593A..65N}.}
\label{Figure11}
\end{figure*}

\begin{figure*}[htbp]
\centering
\includegraphics[height=\textheight, angle=0]{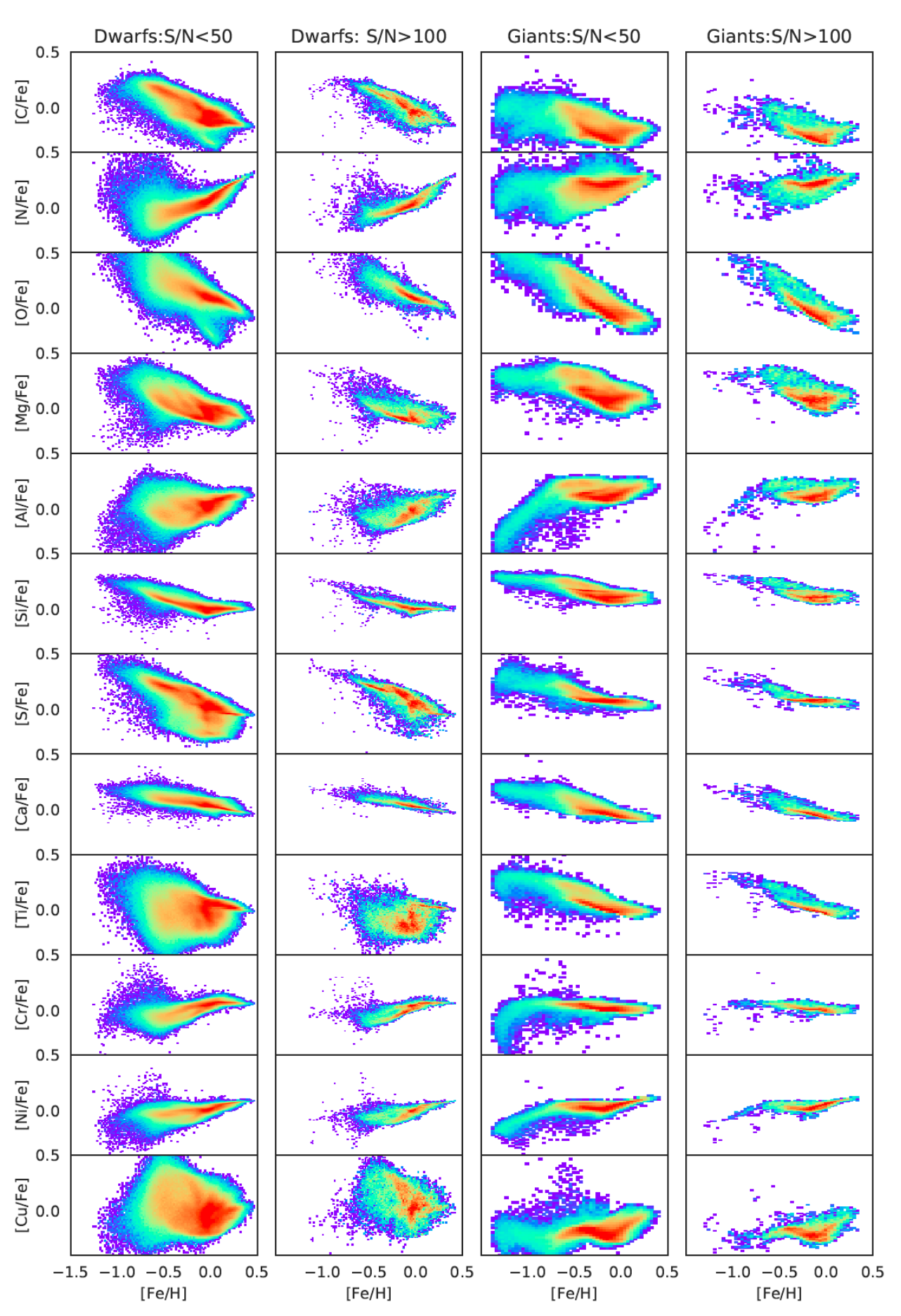}
\caption{Density distribution of 13 elemental abundances for dwarfs and giants at low S/N level ($< 50$) and high S/N level ($> 100$), color-coded by normalized density.}
\label{Figure12}
\end{figure*}

Because the stars with \textit{T}$_{\text{eff}}$ higher than 6500 K show few strong metal lines to measure elemental abundances in LAMOST MRS blue or red bands. We set the element abundances -9999 for these hot stars. Fig.~\ref{Figure10} and Fig.~\ref{Figure11} show density distributions of elemental abundances with respect to [Fe/H] for dwarfs and giants with \textit{T}$_{\text{eff}}$ below 6500 K respectively. Since the MRS sample is dominated by field stars, we expect it to display the known thin/thick disk abundance structure in the [X/Fe] vs [Fe/H] diagram. Alpha-elements [Mg/Fe], [Si/Fe] and [Ti/Fe] show negative correlation and week bimodal structure with respect to [Fe/H] for the giants. The abundance structure for dwarfs and giants at low and high S/N level are displayed in Fig.~\ref{Figure12}. It is apparent that the distribution displayed by the dwarfs and the giants are quite different for a number of elements. Most of the dwarfs concentrates in a tight diagonal sequence which looks inconsistent in position and slope with respect to the giants. These differences are not only a matter of data density in the [X/Fe] vs [Fe/H] plane (which could be attributed to their different spatial sampling, for example, with dwarfs oversampling the thin disk) but also in the structure and position of the sequences displayed by each element. For MRS dwarfs, some systematic are visible at different S/N ranges for most elements. There is not distinct thin/thick sequences visible in the diagram. In addition, there is one sequence with [O/Fe] below 0 in the left panel likely affected by low S/N, which disappears at high S/N level. For MRS giants, the elements of Mg, Al, Si, Ti display the thin/thick disk sequences for both low and high S/N data ($< 50$), while O, Ca and S bimodal sequences become visible only at the high S/N level ($> 100$).

\subsection{Validation}

To ensure the reliability and accuracy of the stellar parameters and chemical abundances obtained with SPCANet model, we employed common stars both from LAMOST-\Rmnum{2} MRS and from some high-resolution observations which have precise stellar parameters, as well as some star clusters to validate the measurement.
\subsubsection{Comparison with other surveys}

\begin{enumerate}

\item{APOGEE} The Apache Point Observatory for Galactic Evolution Experiment~\citep[APOGEE;][]{2015AJ....150..148H, 2017AJ....154...94M} is a median-high resolution (R$\sim$22,500) spectroscopic survey in three near-infrared spectral range ($15700$ to $17500$ \AA). APOGEE DR14~\citep{2018AJ....156..125H,2018AJ....156..126J} published 277,653 spectra, most giants of which have stellar parameters and elements abundances derived by ASPCAP and calibrated using photometric, astroseismology and clusters information. We cross-match our results with APOGEE DR14 and get a subset of 13,184 common stars corresponding 40,122 LAMOST-\Rmnum{2} MRS spectra after setting STARFLAG, ASPCAPFLAG and PARAMFLAG from ASPCAP catalog to ensure the common stars with reliable reference stellar labels.

\item{GALAH} The Galactic Archaeology with HERMES survey~\citep[GALAH;][]{2015MNRAS.449.2604D} make use of a fibre-fed high-resolution (R$\sim$28,000) spectrograph at the 3.9-metre Anglo-Australian Telescope (AAT) to provide multi-object spectra in four spectral ranges ($4713$ to $4903$ \AA , $5648$ to $5873$ \AA , $6478$ to $6737$ \AA , and $7585$ to $7887$ \AA ). The aim of GALAH is to investigate the history of the Galaxy by chemical tagging 30 elements of a million stars. GALAH DR2~\citep{2018MNRAS.478.4513B} has published 342,682 stars with stellar parameters estimated by a multistep approach: the physics-driven \emph{Spectroscopy Made Easy} (SME) followed by the data-driven \emph{The Cannon}, and then 23 elements measured by comparison with MARCS model based synthetic spectra. We cross-match our results with GALAH DR2 and get 396 common stars corresponding 1021 LAMOST-\Rmnum{2} MRS spectra after setting ``flag\_cannon=0" in GALAH parameter catalog.  

\item{RAVE} The RAdial Velocity Experiment~\citep[RAVE;][]{2006AJ....132.1645S} is a medium resolution (R$\sim$7500) spectroscopic survey covering the Ca-triplet spectral region ($8410$ to $8795$ \AA). RAVE DR5~\citep{2017AJ....153...75K} published 520,781 spectra of 457,588 stars, most of which have stellar parameters based on MATISSE~\citep{2006MNRAS.370..141R} and individual abundances for Mg, Al, Si, Ti, Fe, and Ni, relies on a library of equivalent widths~\citep{2011AJ....142..193B}. We cross-match our results with RAVE DR5 and get a subset of 1065 common stars corresponding 3761 LAMOST-\Rmnum{2} MRS spectra after cutting the quality with the flag Algo\_Conv\_K=0 in RAVE parameter catalog.

\end{enumerate}

\begin{figure*}[htbp]
\centering
\includegraphics[width=\textwidth, angle=0]{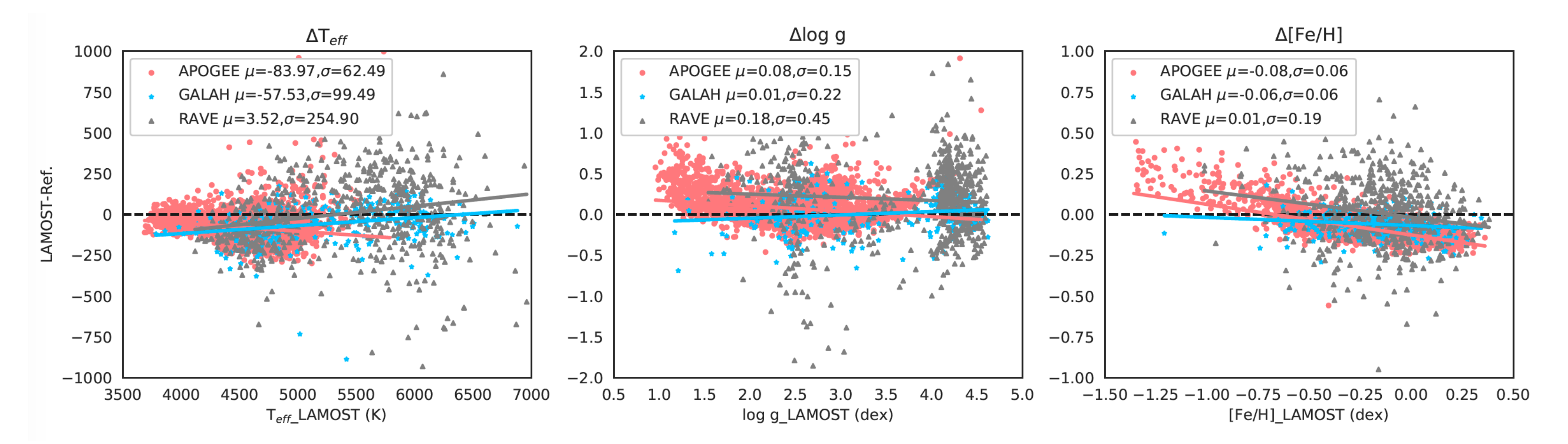}
\caption{Comparison of LAMOST MRS effective temperature in the left panel, surface gravity in the middle panel and [Fe/H] in the right panel, predicted by SPCANet with other surveys' results. The red, blue, grey scatters represent reference values from APOGEE, GALAH and RAVE, respectively. The corresponding lines are their regression lines of their differences with respect to LAMOST values.}
\label{Figure13}
\end{figure*}

\begin{figure*}[htbp]
\centering
\includegraphics[width=\textwidth, angle=0]{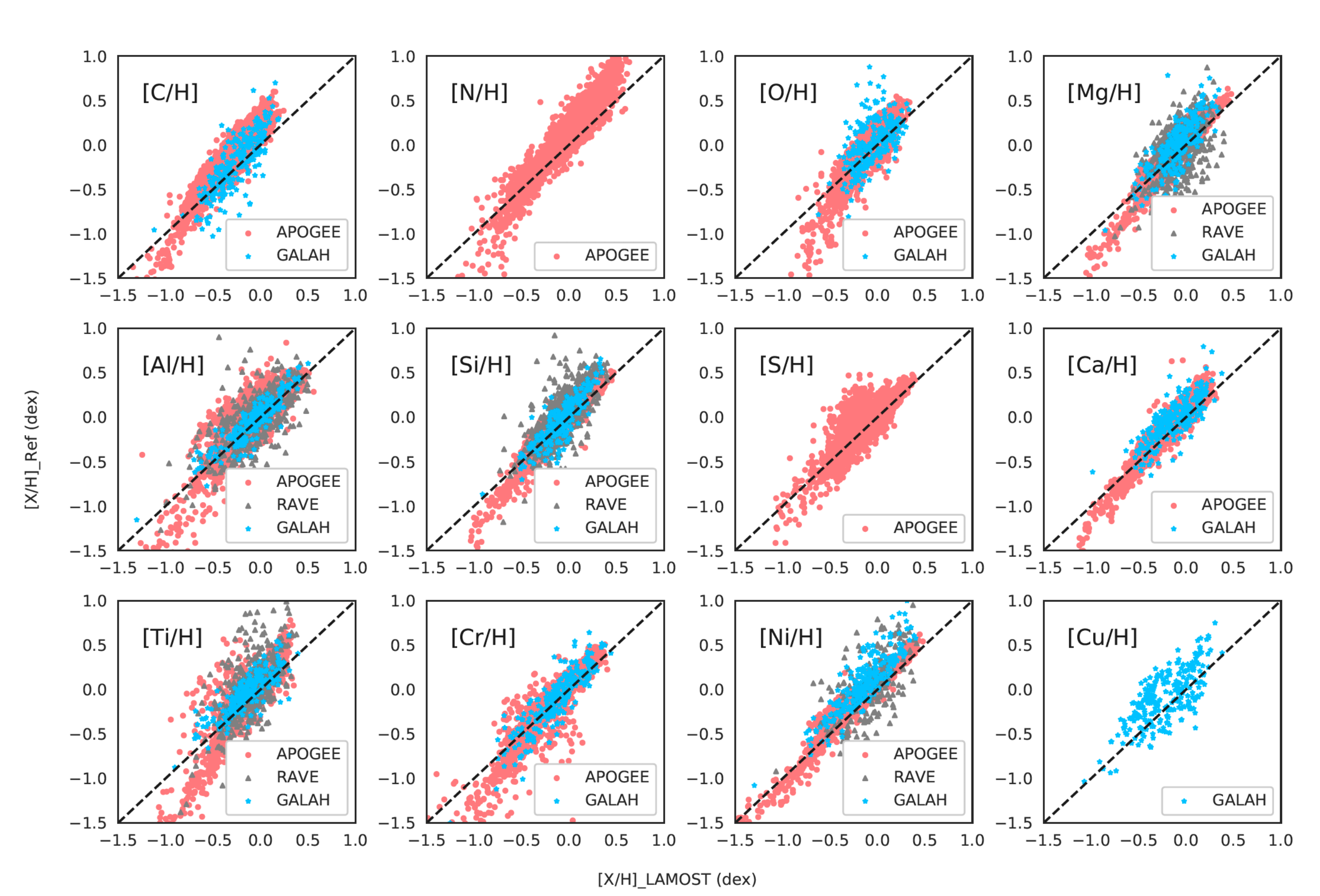}
\caption{Comparison of LAMOST MRS chemical abundances predicted by SPCANet with other surveys' results. The red, blue, grey scatters represent reference values from APOGEE, GALAH and RAVE, respectively. The dotted lines above are one-to-one lines.}
\label{Figure14}
\end{figure*}

\begin{table}[!h]
\caption{Comparison of stellar labels between LAMOST MRS and other surveys.}
\begin{center}
\label{Tab:comparison}
\begin{tabular}{lcccccccccccc}
\hline\noalign{\smallskip}
\multirow{2}{*}{parameters} 			& \multicolumn{2}{c}{APOGEE} & \multicolumn{2}{c}{GALAH} & \multicolumn{2}{c}{RAVE} \\
\cline{2-7} 							& Bias 	&$\sigma$ &Bias &$\sigma$& Bias & $\sigma$\\
\hline\noalign{\smallskip}
$\textit{T}_{\text{eff}}$(K)			& -83.97  & 62.49 &-57.53&99.49  &3.52 & 254.90	\\
log $g$         						&  0.08   & 0.15  &0.01  & 0.22  & 0.18  & 0.45 	\\
$\text{[Fe/H]}$          			&  -0.08  & 0.06  &-0.06 & 0.06  & 0.01  & 0.19  	\\
$\text{[C/H]}$ 						&  -0.20  & 0.11  &-0.04 & 0.19  & -     & -  	\\
$\text{[N/H]}$ 						&  -0.13  & 0.13  &-     & -     & -     & - 		\\
$\text{[O/H]}$ 						&  -0.07  & 0.11  &-0.06 & 0.21  &-  	 & - 	\\
$\text{[Mg/H]}$ 						&  -0.04  & 0.06  &-0.09 & 0.17  & -0.02 & 0.23 	\\
$\text{[Al/H]}$ 						&  -0.04  & 0.12  &-0.04 & 0.10  & -0.02 & 0.20  	\\
$\text{[Si/H]}$ 						&  -0.01  & 0.06  &-0.04 & 0.12  & -0.12 & 0.19 	\\
$\text{[S/H]}$						&  -0.04  & 0.10  & -    & -     & -     & - 		\\
$\text{[Ca/H]}$						&  -0.10  & 0.08  &-0.10 & 0.13  & -	 & -  		\\
$\text{[Ti/H]}$						&  -0.05  & 0.13  &-0.11 & 0.12  & -0.11 & 0.24 	\\
$\text{[Cr/H]}$ 						&  -0.03  & 0.13  &-0.03 & 0.15  & -     & -  	\\
$\text{[Ni/H]}$						&  -0.07  & 0.04  &-0.20 & 0.15  & -0.10 & 0.31  	\\
$\text{[Cu/H]}$ 						&  - 	  & -     &-0.13 & 0.21  & -     & -  	\\
\noalign{\smallskip}\hline
\end{tabular}
\end{center}
\end{table}

Fig.~\ref{Figure13} shows the differences of $\textit{T}_{\text{eff}}$, log $g$ and [Fe/H] between LAMOST-\Rmnum{2} MRS and above three reference sets as a function of LAMOST parameters. LAMOST-\Rmnum{2} MRS temperature appears underestimated comparing with the other three sets. The scatter of the difference between LAMOST and APOGEE is 62.49 K which is less than 99.49 K of GALAH and 257.90 K of RAVE. For log $g$, LAMOST-\Rmnum{2} MRS results are closer to GALAH and APOGEE than to RAVE, and the scatters are smaller with respective to APOGEE than GALAH and RAVE. As we know, the gravities of APOGEE's and RAVE have been calibrated by the asteroseismic gravities or benchmark stars, while gravities of GALAH has not been calibrated. For [Fe/H], the differences between our results and the other surveys' show week systemic trends with small dispersion except for RAVE with a larger dispersion of 0.19 dex. Fig.~\ref{Figure14} shows the comparison between [X/H] of LAMOST MRS and reference sets. The detailed biases and standard deviations are listed in Tab.\ref{Tab:comparison}. On the whole, the scatters are located around the one-to-one line with little dispersion. It should be noted that the biases and dispersion between LAMOST-\Rmnum{2} MRS stellar labels and those from other surveys are contributed by both of our measurements and the reference sets.

\subsubsection{Comparison with the previous literature values}

Open clusters and globular clusters have good chemical consistency, which can be used as good chemical indicators. \citet{2019A&A...622A.191M} provided abundances of light and neutron-capture elements to constrain globular cluster formation by using BACCHUS code analysing APOGEE DR14 spactra. \citet{2018AJ....156..142D} presented analysis of 259 member stars in 19 open clusters from APOGEE DR14 data. \citet{2019A&A...628A..54K} studied the abundances of Fe, Mg, and Ti from medium-resolution spectra of 742 stars in 13 open and globular clusters in the Milky Way. \citet{2017A&A...603A...2M} traced the radial distributions of abundances of elements in the Galactic disc from open clusters and field stars based on Gaia-ESO UVES spectra. \citet{2016AJ....151...49K} analyzed chemical abundances of the 44 members of open cluster M6 based on low-/medium-resolution ESO VLT spectra. \citet{2015A&A...573A..55T} determined C, N and O abundances for stars of Galactic open clusters of the Gaia-ESO survey. Besides, there are also many works focusing on derivation of the chemical composition of field stars. \citet{2011ARep...55..689M} determined abundances of copper, sodium and aluminum of 172 FGK dwarfs from the ELODIE observations. \citep{2014A&A...562A..71B} studied 714 F and G dwarfs and subgiants in the Solar neighborhood and determined their stellar parameters and elemental abundances based on high-resolution spectra. \citet{2016ApJ...833..225Z} presented a study of field stars in the solar neighborhood with non-local thermodynamic equilibrium (NLTE) abundances for 17 chemical elements. \citet{2016A&A...593A..65N} derived very precise abundances of Sc, Mn, Cu and Ba for 21 solar twins and the Sun based on HARPS spectra. We collect the chemical abundance values from above literature for comparison. In total, we get a reference set consist of 3413 stars. Regrettably, most of them have not been visited by LAMOST, we cannot compare their elements with our results one-by-one. Overall trends are shown in [X/Fe] vs. [Fe/H] panels in Fig~\ref{Figure10} for dwarfs and Fig~\ref{Figure11} for giants. We can see that O, Mg, Si, Ca, Cr, Ni show good consistency with the chemical abundances of values from literature. The rest elements could not coincide well, such as C, Al, Ti, Cu because their reference values are widely distributed.

\subsection{LAMOST-\Rmnum{2} MRS catalog of stellar parameters and chemical abundances}

LAMOST-\Rmnum{2} MRS catalog of stellar parameters and chemical abundances contains 1,472,211 spectra. The information published in the on-line catalog contains: the identifier for corresponding star (\emph{starid}), Gaia identifier (Gaia source id), LAMOST spectrum identifier (\emph{medres\_specid}), coordinate information (right ascension (RA), declination (Dec)), signal-to-noise of the spectra (S/N), radial velocities ($\text{RV}_{\text{blue}}$, $\text{RV}_{\text{red}}$) employed from \citet{2019ApJS..244...27W}, effective temperature ($\textit{T}_{\text{eff}}$), surface gravity (log $g$) and elemental abundance ([X/H]) derived by SPCANet. A description of columns of the catalog are shown in Tab~\ref{Tab:description} and the full catalog can be accessed on-line at \url{http://paperdata.china-vo.org/LAMOST/MRS_parameters_elements.csv}.

\begin{table*}[!h]
\caption{Description of the columns of LAMOST MRS stellar parameters and chemical abundances catalog.}
\begin{center}{}
\label{Tab:description}
\begin{tabular}{lll}
\hline\noalign{\smallskip}
\hline\noalign{\smallskip}
Col. & Name  	&  Description \\
\hline\noalign{\smallskip}
1  &$\text{starid}$ & ID for corresponding star based on the R.A. and decl., with the form of ``LAMOST Jdddmmss ddmmss"\\
2  &Gaia$\_{\text{source}\_\text{id}}$ & Gaia source id by crossmatching Gaia DR2\\
3  &$\text{medres\_specid}$ & LAMOST spectral ID, inform of Date-PlateID-SpectrographID-FiberID-MJM-PiplineVersion\\
4  &$\text{medid}_{\text{blue}}$ &  LAMOST spectral ID for the blue part\\
5  &$\text{medid}_{\text{red}}$  &  LAMOST spectral ID for the red part\\
6  &$\text{RA}$ & Right ascension of J2000 ($^{\circ}$) \\
7  &$\text{Dec}$& Declination of J2000 ($^{\circ}$) \\
8  &$\text{S/N}_{\text{blue}}$ & Signal-to-noise of the blue part \\
9  &$\text{S/N}_{\text{red}}$  & Signal-to-noise of the red part \\
10 &$\text{RV}_{\text{blue}}$ & Uncalibrated radial velocity of the blue part ($\text{km}~\text{s}^{-1}$) \\
11 &$\text{RV}_{\text{red}}$  & Uncalibrated radial velocity of the red part ($\text{km}~\text{s}^{-1}$) \\
12 &$\textit{T}_{\text{eff}}$ & Effective temperature (K) \\
13 &log $g$& Surface gravity (dex) \\
14 &[Fe/H] & Iron abundance with respect to hydrogen (dex) \\
15 &[C/H]  & Carbon abundance with respect to hydrogen (dex) \\
16 &[N/H]  & Nitrogen abundance with respect to hydrogen (dex) \\
17 &[O/H]  & Oxygen abundance with respect to hydrogen (dex) \\
18 &[Mg/H] & Magnesium abundance with respect to hydrogen (dex) \\
19 &[Al/H] & Aluminum abundance with respect to hydrogen (dex) \\
20 &[Si/H] & Silicon abundance with respect to hydrogen (dex) \\
21 &[S/H]  & Sulfur abundance with respect to hydrogen (dex) \\
22 &[Ca/H] & Calcium abundance with respect to hydrogen (dex) \\
23 &[Ti/H] & Titanium abundance with respect to hydrogen (dex) \\
24 &[Cr/H] & Cadmium abundance with respect to hydrogen (dex) \\
25 &[Ni/H] & Nickel abundance with respect to hydrogen (dex) \\
26 &[Cu/H] & Copper abundance with respect to hydrogen (dex) \\
27 &Flag & Quality flag: 1 for good, while 0 for bad \\
\noalign{\smallskip}\hline
\end{tabular}
\end{center}
\tablecomments{The full catalog can be accessed on-line.}
\end{table*}

\section{Discussion}
\label{sect:discussion}

We choose from a very precise catalog of stellar parameters and elemental abundances derived by \emph{The Payne} to construct a reference set of stellar labels.  An apposite neural network SPCANet is designed to enable transferring the precise stellar labels to LAMOST-\Rmnum{2} MRS spectra. However, the coverage of the parameters of training set limits the boundary of the SPCANet prediction, which is also the frequent problem that empirical spectral inference always faces. Besides the weak performance of extrapolation using empirical spectral libraries, interpolation does not always success if the training set has maldistribution of samples. From the test set result of SPCANet, we find that the model learns giants much better than dwarfs for the reason that the number of dwarfs in our training sets is far less than giants. To get more precise stellar parameters and elemental abundance with SPCANet for dwarfs of LAMOST-\Rmnum{2} MRS, more and more dwarf stars with known stellar labels need to be observed in medium-resolution mode of LAMOST as benchmark stars for calibration.

A viable method to derive stellar parameters for LAMOST MRS is that mining the physical properties and building a mathematical model behind large amounts of spectra through data-driven methods. An advantage of the data-driven method is that it reduces the additional error introduced during the calculation process because all the input spectra are from the same system. The total error of the final results mostly comes from the contribution of input error. Another viable method is optimizing the similarity measurements of observational spectra with theoretical spectra which based on the stellar atmospheric model and radiation transfer functions. This spectral fitting method depends on perfectness of theoretical model, line spread function (LSF) adjustment and flux calibration, which should be considered carefully when developing a pipeline for stellar parameters and chemical abundances.

\section{Summary}
\label{sect:summary}

We design a new structure of network SPCANet based on a deep-learning method CNN to estimate the stellar atmospheric parameters (\textit{T}$_{\text{eff}}$ and log \textit{g}) and 13 elemental abundances of 1,472,211 spectra from LAMOST-\Rmnum{2} MRS DR7. Then, we utilize some common stars of LAMOST-\Rmnum{2} MRS DR7 and APOGEE-\emph{Payne} to train and test our network. Using the well trained network, we predict stellar parameters and chemical abundances for LAMOST-\Rmnum{2} MRS spectra with S/N$\ge10$, and get the precise of \textit{T}$_{\text{eff}}$, log \textit{g}, [Fe/H] and [X/H] are 119 K, 0.17 dex, 0.06 dex and 0.06$\sim$0.12 dex, except [Cu/H] are 0.19 dex. The results also show good consistency with other surveys such as APOGEE, GALAH and RAVE, as well as the previous literature values, although some small system error exists. 

\section*{Acknowledgement}
This work is supported by the Joint Research Fund in Astronomy (Grant No. U1931209, Grant No.U1631131) under cooperative agreement between the National Natural Science Foundation of China and Chinese Academy of Sciences, China Scholarship Council, the National Natural Science Foundation of China (Grant No.11973060, No.11973022), and Key Research Program of Frontier Sciences, CAS (Grant No.QYZDY-SSW-SLH007). 

Guo Shou Jing Telescope (the Large Sky Area Multi-Object Fiber Spectroscopic Telescope, LAMOST) is a National Major Scientific Project built by the Chinese Academy of Sciences. Funding for the project has been provided by the National Development and Reform Commission. LAMOST is operated and managed by National Astronomical Observatories, Chinese Academy of Sciences.

\software{
		Numpy\citep{oliphant_guide_2006},
		Scipy\citep{jones2001scipy},
		Matplotlib\citep{Hunter:2007},
		Pandas\citep{pythonpandas},
		Keras\citep{chollet2015keras},
		Tensorflow\citep{tensorflow2015-whitepaper},
		Astropy\citep{astropy:2018},
		Spark\citep{Zaharia:2016:ASU:3013530.2934664}
		} 

\bibliography{reference}

\end{document}